\documentclass[aps,pra,twocolumn,groupedaddress,showpacs,10pt]{revtex4-1}
\usepackage{hyperref}
\usepackage{bm}
\usepackage{graphicx}

\begin{document}


\title{Assessing thermalization and estimating the Hamiltonian with output data only}


\author{Jochen Rau}
\email[]{jochen.rau@q-info.org}
\homepage[]{www.q-info.org}
\affiliation{Institut f\"ur Theoretische Physik,
Johann Wolfgang Goethe-Universit\"at,
Max-von-Laue-Str. 1, 
60438 Frankfurt am Main,
Germany}


\date{\today}

\begin{abstract}
I consider the generic situation where a finite number of identical test systems in varying (possibly unknown) initial states are subjected independently to the same unknown process.
I show how one can infer from the output data alone
whether or not the process in question induces thermalization, and if so,
which constants of the motion characterize the final equilibrium states.
In case thermalization does occur and there is no evidence for constants of the motion other than energy,
I further show how the same output data can be used to estimate the test systems' effective Hamiltonian.
For both inference tasks I devise a statistical framework inspired by the generic  techniques of factor and principal component analysis.
I illustrate its use in the simple example of qubits.
\end{abstract}


\maketitle

\section{\label{intro}Introduction}

Controversies over the apparent dichotomy between microscopic reversibility and macroscopic irreversibility are as old as statistical mechanics itself and continue to the present day,
as exemplified by the popular Ref. \cite{lebowitz:boltzmann} and the ensuing vivid debate \cite{barnumetal:letters}.
Broadly speaking, the issue can be tackled 
``bottom up'' or ``top down''.
The bottom up approach, which has been pursued by the majority of researchers,
involves specifying (or at least imposing constraints on)
some microscopic Hamiltonian and subsequently studying the evolution of those degrees of freedom that are deemed ``macroscopic'', ``accessible'' or otherwise ``relevant'' to the problem at hand.
This line of research has of late enjoyed cross-fertilization with topical areas such as 
nanoscale thermodynamics \cite{allahverdyan:nano,allahverdyan:work,janzing:molecular,PhysRevLett.104.090602}, 
quantum many-body physics \cite{rigol:gge,PhysRevLett.103.100403,PhysRevLett.105.260402,PhysRevLett.106.040401,RevModPhys.83.863,trotzky:1dbose} 
and quantum information \cite{PhysRevLett.96.050403,popescu:entanglement,brandao:entanglement+2ndlaw,springerlink:10.1007/s00220-010-1003-1},
leading to some powerful new results.
They confirm that the eventual thermalization of a quantum system
is a universal phenomenon which holds true for virtually all Hamiltonians and sensible choices for the relevant degrees of freedom \cite{PhysRevLett.101.190403,PhysRevE.79.061103,1367-2630-13-5-053009,riera:thermalization}.
The rather generic assumptions that are needed amount to 
(i)
excluding the special case of isolated systems with highly regular, completely integrable dynamics;
and 
(ii)
introducing some form of coarse graining,
such as 
limiting the resolution of realistic preparation and measurement devices \cite{PhysRevLett.101.190403}
or 
tracing out the degrees of freedom of a bath \cite{PhysRevE.79.061103}.
Coarse graining entails that information about the microstate is siphoned off from the retained to the discarded degrees of freedom.
This leakage becomes irreversible whenever the dynamics of the latter is sufficiently fast and irregular,
leading to an effective memory loss on time scales much shorter than those pertaining to the evolution of the relevant degrees of freedom \cite{rau:physrep}.

In contrast, 
the lesser-known top down approach, pioneered by Jaynes for classical statistical mechanics \cite{jaynes:2ndlaw} and subsequently generalized \cite{rau:convergence}, refrains from considering any specifics of the underlying microscopic dynamics and instead derives macroscopic irreversibility from the very basic requirement -- essential to the scientific method -- that macroscopic experiments be reproducible.
The central argument is very simple:
An experiment is reproducible if its initial preparation uniquely determines its final outcome;
i.e., if merely on the basis of their initial values one can predict with  certainty the final values of the relevant degrees of freedom.
Since a prediction cannot possibly contain more information than the data on which it is based,
the final values of the relevant degrees of freedom cannot carry more information than do their initial values.
So in the course of a reproducible experiment 
the amount of {missing} information about the system's microstate, 
and hence the entropy,
can only increase, Q.E.D.
There are other top down approaches which are similar in spirit,
yet which rather than from ``reproducibility'' 
start from different primitives like ``adiabatic accessibility'' \cite{Lieb19991}.

Reversing the top down logic, 
violations of the second law may well occur;
but such violations are never reproducible,
and with increasing system size, become exceedingly unlikely.  
Experiments that purport to violate the second law in a reproducible fashion
must presuppose the preparation of 
some special (say, highly correlated) initial state,
or else some peculiar prior history of the system (such as in the classic example of spin echoes \cite{hahn:spinechoes}).
The apparent systematic violation of the second law then stems from the fact that
the experimenter actually controls degrees of freedom other than the supposedly relevant ones,
either directly in the present or through specific interventions in the past.

Despite their seemingly different outlooks
the bottom up and top down approaches both revolve around the pivotal issue of memory loss.
They either show (bottom up) or simply postulate (top down) that in realistic experiments the relevant degrees' remote history has no influence on their future evolution, and thus can be safely disregarded.
This intimate connection between irreversibility and memory loss is captured succinctly in Landauer's principle \cite{landauer:principle}, which has spawned another highly interesting line of research \cite{springerlink:10.1007/BF02084158,PhysRevLett.100.080403,PhysRevLett.102.250602,PhysRevE.83.030102,delrio:negentropy}.

In the present paper I wish to add yet another, and rather practical, perspective on the issue of thermalization.
When a novel quantum system is fabricated and investigated in the laboratory for the first time,
its precise dynamics and possibly even its constants of the motion are not known in advance. 
(Of course, there is generally some theoretical expectation;
but whether this will be confirmed or refuted by actual measurements is not \textit{a priori} clear.)
A particular experiment might then be aimed at assessing whether or not a certain process leads to thermalization;
and if so, 
which set of thermodynamic variables characterizes the final equilibrium state.
Operationally, one might do this by assembling multiple samples,
each consisting of identically prepared copies of the system.
Each sample is prepared in a different initial state and subjected to the process in question.
If thermalization does occur, subsequent quantum-state tomography \cite{book:stateestimation} on all samples will reveal that, modulo random fluctuations, their respective final states are distributed on some low-dimensional submanifold of state space. 
This submanifold is composed of states of the Gibbs form 
$\rho\propto\exp(-\sum_a \lambda^a G_a)$, with the observables $\{G_a\}$ being the constants of the motion.
Their expectation values or the associated Lagrange parameters, respectively,
then constitute the appropriate set of thermodynamic variables. 
This approach to assessing thermalization is based on output data only
and does not require tight control over the initial states of the various samples.

Yet in a real-world setting, the system in question might be difficult to manufacture,
and the above idealized procedure difficult to execute.
Specifically, it might only be possible to prepare a small number of samples,
which in turn are small in size.
As a consequence, there will be just a few data points in state space, which moreover have non-negligible error bars. 
Reconstructing the Gibbs manifold and hence the constants of the motion
on the basis of such imperfect measurement data
then becomes a nontrivial statistical inference task.
In purely statistical terms, this is a situation where noisy data in some high-dimensional space (the tomographic images in state space) are presumed to be explained by a small number of latent variables (the expectation values of the constants of the motion), effectively reducing the dimensionality of the data. 
In such a generic setting, the task is to infer the optimal dimension and orientation of the lower-dimensional latent space. 
Problems of this type can be tackled with a variety of statistical techniques such as factor analysis or principal component analysis \cite{roweis:em,roweis:unify,bishop:bayesian_pca,bishop:variational_pca,RSSB:RSSB196,tipping:mixtures,roweis:nonlinear,gorban:principalgraphs}. 
In the present paper, I shall build on these generic techniques to develop a statistical framework  
tailored to the relevant task of assessing whether or not thermalization has occurred, and if so, inferring the most plausible set of constants of the motion.

Whenever the above statistical analysis suggests that thermalization has indeed occurred and there is one constant of the motion only,
this single constant of the motion is by default the Hamiltonian.
The same statistical framework can then be used to 
estimate that Hamiltonian.
This estimation procedure is based on studying thermal properties rather than time evolution;
and it uses only output rather than input-output data.
Therefore, it is very different in its approach from the usual quantum-process tomography \cite{nielsen:book,PhysRevLett.78.390,chuang:processtomography,PhysRevA.63.054104,PhysRevLett.86.4195,10.1063/1.1518554,PhysRevLett.90.193601,PhysRevA.72.022106,PhysRevA.78.032118,PhysRevA.77.032322}
and
Hamiltonian tomography \cite{PhysRevA.77.042320,PhysRevA.80.022333,springerlink:10.1134/S1054660X10090434,schirmer:isccsp,PhysRevA.84.012107}. 
As the second key result of the present paper,
I shall lay out this ``thermal'' estimation procedure for the Hamiltonian
and illustrate its use in a simple example.

The remainder of the paper is organized as follows.
In Sec. \ref{theory},
I will present the general statistical framework for assessing thermalization along with the key approximations made.
In Sec. \ref{hamest},
I will turn to the rather common case where the Hamiltonian is the sole constant of the motion,
and explain how in this case one can infer the most plausible Hamiltonian from the data.
In Sec. \ref{qubit},
I will put the general framework to use in the simple example of qubits,
both to assess their thermalization and to estimate the pertinent Hamiltonian. 
Finally, in Sec. \ref{conclusions},
I will conclude with a brief discussion.

\section{\label{theory}Assessing thermalization}

Let $R$ denote the number of distinct samples
and $N_i, i=1\ldots R$, the size of the $i$-th sample.
After the samples have undergone the process in question
they are all subjected to quantum-state tomography,
which may or may not be informationally complete.
Let $\{F_b\}$ denote the set of observables whose totals are ascertained in a tomographic experiment
(by performing measurements on each member of the sample and adding up the results,
or via global measurements on the entire sample),
and $\{f^i_b\}$ the associated sample means gleaned from the $i$-th sample.
Finally, let 
the quantum state $\sigma$ denote a possible prior bias as to the samples' final state \cite{rau:evidence};
in case of complete prior ignorance, this is simply taken to be the totally mixed state.

The hypothesis to be tested is whether or not the totality of experimental data $D\equiv \{f^i_b\}$ can be explained by the expectation values of some smaller set of observables $\{G_a\}$,
the presumed constants of the motion.
Associated 
with these presumed constants of the motion 
and 
with the measured observables 
are subspaces
${\cal G}:=\mbox{span}\{1,G_a\}$
and
${\cal F}:=\mbox{span}\{1,F_b\}$
of the space of observables
(with $1$ being the unit operator),
termed respectively the ``theoretical'' and ``experimental'' level of description \cite{PhysRevA.84.012101}.
For the former to have any explanatory value, it must be 
$\dim{\cal G}<\dim{\cal F}$.

The plausibility of the theoretical hypothesis
is encoded in
the posterior probability of the  level of description ${\cal G}$,
given the data $D$ and prior bias $\sigma$.
By Bayes' rule \cite{schack:bayesrule},
this probability of interest is given by
\begin{equation}
	\mbox{prob}({\cal G}|D,\{N_i\},{\cal F};\sigma)
	\propto
	\mbox{prob}({\cal G})
	\mbox{prob}(D|\{N_i\},{\cal F};\sigma,{\cal G})
	.
\end{equation}
Whenever the prior $\mbox{prob}({\cal G})$ is sufficiently non-committal, 
the right hand side will be dominated by the likelihood function.
As the various runs of the experiment are independent,
the latter can be factorized:
\begin{equation}
	\mbox{prob}(D|\{N_i\},{\cal F};\sigma,{\cal G})
	=
	\prod_{i=1}^R \mbox{prob}(D_i|N_i,{\cal F};\sigma,{\cal G})
	,
\end{equation}
with the data $D_i$ pertaining to the $i$-th sample. 
And finally, according to the theoretical hypothesis, each individual factor can be marginalized:
\begin{eqnarray}
	\lefteqn{
	\mbox{prob}(D_i|N_i,{\cal F};\sigma,{\cal G})
	}
	\nonumber \\
	&=&
	\int_{\pi^\sigma_{\cal G}({\cal S})} d\omega\ 
	\mbox{prob}(D_i|N_i,\omega,{\cal F}) \mbox{prob}(\omega|\sigma,{\cal G})
	,
\label{marginalize}
\end{eqnarray}
where
the integration ranges not over the complete state space ${\cal S}$
but over the Gibbs manifold 
$\pi^\sigma_{\cal G}({\cal S})$ associated with the theoretical level of description ${\cal G}$ and reference state $\sigma$ \cite{PhysRevA.84.012101}.
This Gibbs manifold is composed of states of the generalized Gibbs form
\begin{equation}
	\omega \propto \exp\left[ (\ln\sigma-\langle\ln\sigma\rangle_\sigma)-\sum_a \lambda^{a}_i G_a \right]
	,
\label{gibbs}
\end{equation}
which minimize the relative entropy with respect to $\sigma$ under given constraints for the expectation values of $\{G_a\}$ \cite{ruskai:minrent,olivares+paris}.

For reasonably large sample sizes and a near optimal measurement setup the first factor (likelihood) in the integrand of Eq. (\ref{marginalize})
can be approximated with the help of the quantum Stein lemma \cite{hiai+petz,ogawa+nagaoka,vedral:rmp,brandao+plenio},
\begin{equation}
	\mbox{prob}(D_i|N_i,\omega,{\cal F})
	\propto
	\exp[-N_i S(\mu_i\|\omega)]
	.
\label{entropic_likelihood}
\end{equation}
(Else the quantum Stein lemma provides only a lower bound.)
Here $\mu_i\in\pi^\omega_{\cal F}({\cal S})$ has the generalized Gibbs form (\ref{gibbs}) 
with $\{G_a\}$ replaced by $\{F_b\}$ and reference state $\omega$ rather than $\sigma$,
and with the Lagrange parameters $\{\lambda^{b}_i\}$ adjusted such that $\langle F_b\rangle_{\mu_i}=f^i_b$ for all $b$.
For both conceptual and practical reasons
I shall model the second factor (prior) in the integrand as an entropic distribution, too,
\begin{equation}
	\mbox{prob}(\omega|\sigma,{\cal G})
	\propto
	\exp[-\alpha S(\omega\|\sigma)] 
	,
\label{entropic_ansatz}
\end{equation}
with a factor of proportionality that does not depend on $\omega$ \cite{PhysRevA.84.012101}.
This ansatz contains an unknown hyperparameter $\alpha>0$,
whose most likely value will be estimated later via the evidence procedure.

I assume that the theoretical level of description is a proper subspace of the experimental level,
${\cal G}\subset {\cal F}$,
so that $\pi^\omega_{\cal F}({\cal S})=\pi^\sigma_{\cal F}({\cal S})$.
The Gibbs manifold $\pi^\sigma_{\cal G}({\cal S})$,
which contains the theoretical models $\omega$ and the reference state $\sigma$,
is then a proper submanifold of $\pi^\sigma_{\cal F}({\cal S})$ which contains the tomographic images $\{\mu_i\}$.
Each tomographic image $\mu_i$ has a unique projection 
$\pi_i:=\pi^\sigma_{\cal G}(\mu_i)$ on the submanifold $\pi^\sigma_{\cal G}({\cal S})$,
where $\pi^\sigma_{\cal G}$ is the coarse graining operation that maps arbitrary states to
Gibbs states on $\pi^\sigma_{\cal G}({\cal S})$, thereby preserving the expectation values of
the relevant observables $\{G_a\}$.
Also on the submanifold $\pi^\sigma_{\cal G}({\cal S})$,
between the projection $\pi_i$ and the reference state $\sigma$, lies 
the 
interpolated state \cite{mackay:interpolation,PhysRevA.84.012101}
\begin{equation}
	{\rho_i}
	:\propto
	\exp\left[(1-x_i) \ln\pi_i + x_i \ln\sigma \right]
\label{interpolation}
\end{equation}
with $x_i:=\alpha/(\alpha+N_i)$;
its Lagrange parameters are the weighted average of those of $\pi_i$ and $\sigma$,
with respective weights $N_i$ and $\alpha$.
Finally,
for both the tomographic images $\{\mu_i\}$ and their projections $\{\pi_i\}$
one defines respective center-of-mass states
\begin{equation}
	\tilde{\mu}:\propto \exp\left[\sum_{i=1}^R w_i \ln\mu_i\right]
	\ ,\ 
	\tilde{\pi}:\propto \exp\left[\sum_{i=1}^R w_i \ln\pi_i\right]
\end{equation}
with $w_i:=N_i/\sum_j N_j$,
which lie on $\pi^\sigma_{\cal F}({\cal S})$ and $\pi^\sigma_{\cal G}({\cal S})$,
respectively,
and which are obtained by taking the weighted average over all samples of the respective Lagrange parameters.

\begin{figure}[htbp]
\begin{center}
\includegraphics[width=8.5cm]{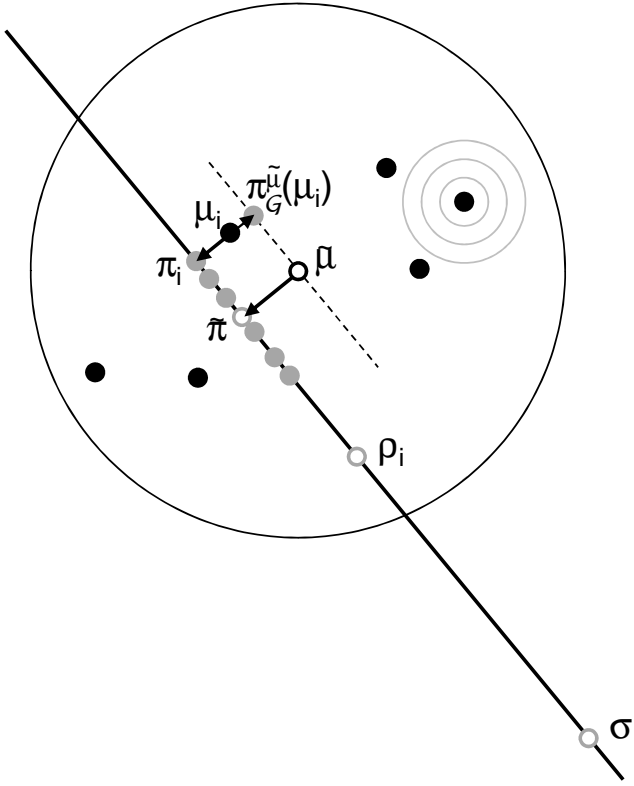}
\end{center}
\caption{
States on the manifold $\pi^\sigma_{\cal F}({\cal S})$.
Black dots indicate the tomographic images $\{\mu_i\}$ associated with data garnered from different samples,
and the small black circle their center of mass $\tilde{\mu}$.
The straight lines are the reduced Gibbs manifolds $\pi^\sigma_{\cal G}({\cal S})$ (solid line)
and $\pi^{\tilde{\mu}}_{\cal G}({\cal S})$ (dashed line), respectively.
Gray dots or circles denote states on either of these reduced Gibbs manifolds.
In particular,
the gray dots are  
obtained by applying the coarse graining $\pi^\sigma_{\cal G}$ or $\pi^{\bar{\mu}}_{\cal G}$, respectively,
to the tomographic images.
(For simplicity, not all coarse grainings are shown.)
The state $\rho_i$ is the interpolation (\ref{interpolation}) between the coarse grained image 
$\pi_i$ and the reference state $\sigma$.
All states inside the big circle are assumed to be sufficiently close to each other 
to warrant the Gaussian approximation for their relative entropies;
the only state that might lie outside this Gaussian region is the reference state $\sigma$.
The gray concentric circles around one of the tomographic images indicate an exemplary likelihood function (\ref{entropic_likelihood}).
It has a width of order $1/\sqrt{N_i}$,
which is assumed to lie inside the Gaussian region.
}
\label{gauss}
\end{figure}

For nearby states on the manifold $\pi^\sigma_{\cal F}({\cal S})$
the relative entropy is approximately quadratic in their coordinate differentials,
\begin{equation}
	S(\mu\|\mu')
	\approx
	(1/2)
	\sum_{ab} (C^{-1})^{ab} \delta f_a \delta f_b
	.
\end{equation}
Here $C$ denotes the correlation matrix
\begin{equation}
	C_{ab}(\rho):=\langle \delta F_a;\delta F_b\rangle_\rho
\label{correlation}
\end{equation}
with $\delta F_b:=F_b-\langle F_b\rangle_\rho$ and canonical correlation function
\begin{equation}
	\langle X;Y\rangle_\rho:=\int_0^1 d\nu\,\mbox{tr}({\rho}^\nu X {\rho}^{1-\nu} Y)
	.
\end{equation}
The correlation matrix varies little between $\mu$ and $\mu'$,
and so to lowest order, can be evaluated in either of the two states
or in any other state $\rho$ in their vicinity.
In the following
I shall assume that the tomographic images $\{\mu_i\}$,
their projections $\{\pi_i\}$, as well as their respective centers of mass
$\tilde{\mu}$ and $\tilde{\pi}$ all lie inside a region 
in which the above quadratic (``Gaussian'') approximation is warranted,
with the correlation matrix evaluated in the center of mass $\tilde{\mu}$,
$C=C(\tilde{\mu})$.
This presupposes that for all samples the presumed constants of the motion take values within a sufficiently narrow range.
Moreover,
I shall assume that 
the sample sizes $\{N_i\}$ are sufficiently large compared to $\alpha$ so that 
the interpolated states $\{\rho_i\}$, too, lie inside this region.
And finally, I assume that the sample sizes are also large enough in absolute terms
to render the likelihood function (\ref{entropic_likelihood}) 
largely concentrated inside the Gaussian region.
The reference state $\sigma$, on the other hand, need not necessarily be inside the Gaussian region (Fig. \ref{gauss}).

The confinement of all pertinent states (with the exception of the reference state) to a Gaussian region
entails a number of simplifications:
Relative entropies become approximately symmetric,
$S(\mu\|\mu')\approx S(\mu'\|\mu)$;
for the interpolated states $\{\rho_i\}$, it is
\begin{equation}
	(1-x_i) S(\rho_i\|\pi_i) + x_i S(\rho_i\|\sigma)
	\approx
	x_i(1-x_i) S(\pi_i\|\sigma)
\end{equation}
up to corrections of order $O((\alpha/N_i)^2)$ that account for the possible non-Gaussianity of $S(\pi_i\|\sigma)$;
the centers of mass $\tilde{\mu}$ and $\tilde{\pi}$ coincide approximately with the ordinary mixtures
\begin{equation}
	\bar{\mu}:=\sum_{i=1}^R w_i \mu_i
	\ ,\ 
	\bar{\pi}:=\sum_{i=1}^R w_i \pi_i
	,
\label{mixture}
\end{equation}
respectively;
and
the coarse graining map is approximately linear, so
$\bar{\pi}
	\approx 
	\pi^\sigma_{\cal G}(\bar{\mu})$
	.

Using these approximations, as well as the
(exact) law of Pythagoras \cite{petz:book}
\begin{equation}
	S(\mu_i\|\omega) 
	= 
	S(\mu_i\|\pi_i) 
	+ S(\pi_i\|\omega)
\label{pythagoras}
\end{equation}
for all $\omega\in\pi^\sigma_{\cal G}({\cal S})$
and the (exact)
mixing rules
\begin{eqnarray}
	\lefteqn{
	(1-x_i) S(\omega\|\pi_i) + x_i S(\omega\|\sigma)
	}
	\nonumber \\
	&=&
	(1-x_i) S(\rho_i\|\pi_i) + x_i S(\rho_i\|\sigma)
	+ S(\omega\|\rho_i)
\end{eqnarray}
and
\begin{equation}
	\sum_{i=1}^R w_i S(\pi_i\|\sigma)
	=
	\sum_{i=1}^R w_i S(\pi_i\|\bar{\pi}) + S(\bar{\pi}\|\sigma)
	,
\end{equation}
one obtains the log-likelihood
\begin{eqnarray}
	\lefteqn{
	\ln \mbox{prob}(D|\{N_i\},{\cal F};\sigma,{\cal G})
	}
	\nonumber \\
	&\approx&
	\sum_{i=1}^R N_i [S(\pi_i\|\bar{\pi}) - S(\bar{\mu}\|\bar{\pi})]
	- \frac{p\Lambda}{2} 
	\nonumber \\
	&&
	- \sum_{i=1}^R x_i N_i [S(\pi_i\|\bar{\pi}) + S(\bar{\pi}\|\sigma)]
	+ \frac{p}{2} \sum_{i=1}^R \ln (x_i N_i)
\label{loglike}
\end{eqnarray}
with $p:=\dim\pi^\sigma_{\cal G}({\cal S})$ and $\Lambda:=\sum_i \ln N_i$,
modulo a small correction term that accounts for the possible non-Gaussianity of $S(\bar{\pi}\|\sigma)$
and varies only weakly with $\alpha$,
and
modulo additive constants that do not depend on $\alpha$, $\sigma$ or ${\cal G}$.
Since $x_i N_i = (1-x_i)\alpha\le\alpha$,
the terms in the last row of Eq. (\ref{loglike}) do not scale with sample size (at fixed $\alpha$)
and so become negligible in the regime $N_i\gg\alpha$. 
The log-likelihood then approaches 
(again modulo additive constants that do not depend on $\sigma$ or ${\cal G}$) 
the asymptotic result
\begin{equation}
	L({\cal G})
	:=
	\sum_{i=1}^R N_i [S(\pi_i\|\bar{\pi}) - S(\bar{\mu}\|\bar{\pi})]
	- \frac{p\Lambda}{2} 
	.
\label{asymptotic}
\end{equation}
This asymptotic log-likelihood is the central quantity which I will use for my subsequent analysis.

Strictly speaking, one has yet to check that it is consistent to assume that $\alpha$ stays constant when taking the limit $N_i\to\infty$;
i.e., that the most likely value of $\alpha$ does not itself scale with sample size.
In order to determine this most likely value,
I follow the prescription of the evidence procedure \cite{rau:evidence}.
I consider the log-likelihood (\ref{loglike}) and seek its maximum as a function of $\alpha$.
Setting its derivative with respect to $\alpha$ equal to zero yields the extremum condition
\begin{equation}
	\sum_{i=1}^R (1-x_i) N_i
	\left\{
		x_i [S(\pi_i\|\bar{\pi}) + S(\bar{\pi}\|\sigma)] - \frac{p}{2N_i}
	\right\}
	=
	0
	.
\end{equation}
(This maximum likelihood condition generalizes an earlier result for experiments on a single sample  \cite{rau:evidence,PhysRevA.84.012101}.)
In the asymptotic limit $N_i\to\infty$ (at fixed relative entropies), the maximum likelihood estimates for the $\{x_i\}$ must scale as the inverse sample size;
and so indeed, $\alpha=x_i N_i/(1-x_i)$
must \textit{not} scale with sample size.
This conclusion about $\alpha$ is robust as long as
\begin{equation}
	-\alpha^2 \frac{\partial^2}{\partial \alpha^2}
	\ln \mbox{prob}(D|\{N_i\},{\cal F};\sigma,{\cal G})
	\gg
	1
	.
\end{equation}
In the relevant regime $N_i\gg\alpha$ the left hand side of this condition is approximately $pR/2$,
so one has good accuracy  
whenever the number of samples is sufficiently large, $R\gg 1$.

The asymptotic log-likelihood (\ref{asymptotic}) is the difference of two sums,
reflecting a trade-off that is typical for model selection \cite{sivia:modelselection}.
The first sum gets bigger as the theoretical level of description becomes more detailed and yields a better fit with the data;
in fact, it is maximal for the largest possible level of description, ${\cal G}={\cal F}$.
The sum which is subtracted from this, on the other hand, being proportional to the Gibbs manifold dimension,
penalizes excessive detail;
it embodies ``Occam's razor''.
Therefore, finding the most plausible level of description and hence the constants of the motion
always involves a trade-off between goodness of fit and simplicity.

In case the reference state $\sigma$ is not given \textit{a priori} but is itself a variable to be inferred,
one must consider the asymptotic log-likelihood (\ref{asymptotic}) also as a function of $\sigma$.
The log-likelihood attains its maximum for any 
$\sigma\in\pi^{\bar{\mu}}_{\cal G}({\cal S})$;
then the relative entropy $S(\bar{\mu}\|\bar{\pi})$ vanishes.
Using such a maximum likelihood estimate for $\sigma$,
and assuming further that
the dimension $p$ of the Gibbs manifold is fixed from the outset,
the remaining optimization of (the orientation of) ${\cal G}$ reduces to maximizing the weighted average of the relative entropies $\{S(\pi^{\bar{\mu}}_{\cal G}(\mu_i)\|\bar{\mu})\}$.
In the Gaussian regime,
this is tantamount to the optimization task known in statistics as
{``principal component analysis''} \cite{roweis:em,roweis:unify,bishop:bayesian_pca,bishop:variational_pca,RSSB:RSSB196,tipping:mixtures,roweis:nonlinear,gorban:principalgraphs}.

Now I turn to the general case in which there is an arbitrary given reference state, 
and where both the dimension and the orientation of the explanatory level of description 
are  to be inferred.
Suppose there are two rival proposals for the level of description, ${\cal G}$ and ${\cal H}$,
where the latter is more detailed than the former (and both are contained in the experimental level of description),
${\cal G}\subset{\cal H}\subset{\cal F}$.
The associated Gibbs manifolds $\pi^\sigma_{\cal G}({\cal S})$ and $\pi^\sigma_{\cal H}({\cal S})$
have respective manifold dimensions $p$ and $p+s$.
As discussed earlier,
the choice between the two proposals will involve a trade-off between goodness of fit (favoring ${\cal H}$)
and simplicity (favoring ${\cal G}$).
Using the fact that within the Gaussian region
the relative entropy of two coarse grained states
is approximately invariant under a change of reference state $\sigma\to{\bar{\mu}}$,
\begin{equation}
	S(\pi_i\|\bar{\pi})
	\approx
	S(\pi^{\bar{\mu}}_{\cal G}(\mu_i)\|\bar{\mu})
\label{shiftreference}
\end{equation}
(and likewise for ${\cal H}$),
the difference of the asymptotic log-likelihoods can be written as
\begin{eqnarray}
	\lefteqn{L({\cal H}) - L({\cal G})}
	\nonumber \\
	&\sim&
	\sum_{i=1}^R N_i 
	[
	S(\pi^{\bar{\mu}}_{\cal H}(\mu_i)\|\pi^{\bar{\mu}}_{\cal G}(\mu_i))
	+ 
	S(\pi^\sigma_{\cal H}(\bar{\mu})\|\pi^\sigma_{\cal G}(\bar{\mu}))
	]
	-
	\frac{s\Lambda}{2} 
	.
	\nonumber \\
\label{criterion}
\end{eqnarray}
If this difference is positive, the more detailed level of description ${\cal H}$ is called for;
if it is negative, one better stick to the simpler model ${\cal G}$.
This criterion extends an earlier result obtained in Ref. \cite{PhysRevA.84.012101} for experiments on a single sample.

Finding the optimal level of description, and hence the most plausible set of constants of the motion,
can now proceed in two ways:
either directly,
by maximizing the asymptotic log-likelihood (\ref{asymptotic}) as a function of ${\cal G}$;
or indirectly (and usually more feasible in practice),
by formulating various hypotheses about the level of description and then comparing them by means of the difference criterion (\ref{criterion}).
If the optimal ${\cal G}$ is spanned by only one or very few observables (aside from the unit operator),
this indicates that thermalization has indeed occurred.

The reconstruction of the appropriate level of description precedes the reconstruction of the quantum state of any individual system.
The former requires data from the totality of all samples.
Once the reconstruction of the level of description has succeeded,
one may  take this level as a given and turn to reconstructing
the Gibbs state of an {{individual}} system,
based on data from the pertinent sample only,
by means of well-known state estimation techniques \cite{PhysRevA.84.012101}.

\section{\label{hamest}Hamiltonian estimation}

Whenever the above statistical analysis reveals or it is posited from the outset
that there is only one constant of the motion,
this is by default the Hamiltonian.
The Gibbs manifold is then made up of canonical states $\rho\propto\exp(-\beta H)$,
with Hamiltonian $H$ and inverse temperature $\beta$.
(For a non-uniform reference state there is an additional term $(\ln\sigma-\langle\ln\sigma\rangle_\sigma)$ in the exponent.)
Strictly speaking, in case the $\{F_b\}$ are not informationally complete,
$H$ is not the full Hamiltonian but the \textit{effective} Hamiltonian pertaining to the measured degrees of freedom.
Since the latter usually coincide with the slow degrees of freedom,
$H$ is then an effective low-energy Hamiltonian.
If the Hamiltonian is not known in advance, it must be estimated from the data.
In this section, I shall lay out the appropriate estimation procedure.

Let the Hamiltonian be parametrized by some set of parameters $\xi\equiv\{\xi^b\}$.
Then so are the coarse grained states 
\begin{equation}
	\pi_i(\xi)
	=
	Z(\beta_i,\xi)^{-1}
	\exp[(\ln\sigma-\langle\ln\sigma\rangle_\sigma) - \beta_i H(\xi)]
	,
\end{equation}
with arbitrary reference state $\sigma$,
where the partition function
\begin{equation}
	Z(\beta_i,\xi)
	:=
	\mbox{tr}\{
	\exp[(\ln\sigma-\langle\ln\sigma\rangle_\sigma) - \beta_i H(\xi)]
	\}
\end{equation}
ensures state normalization,
and the inverse temperature $\beta_i$ is adjusted such that 
$\langle H(\xi)\rangle_{\pi_i(\xi)}=\langle H(\xi)\rangle_{\mu_i}=:U_i$.
Their weighted average $\bar{\pi}(\xi)$,
equally parametrized by $\xi$,
has (in the Gaussian approximation) the same canonical form, with
inverse temperature
$\bar{\beta}\approx\sum_i w_i \beta_i$
and internal energy
$\bar{U}=\sum_i w_i U_i$.
The asymptotic log-likelihood (\ref{asymptotic}) thus becomes a function of $\xi$.
It attains its maximum when
\begin{equation}
	{\partial_\xi} \sum_{i=1}^R w_i S(\pi_i(\xi)\|\bar{\pi}(\xi))
	=
	{\partial_\xi S(\bar{\mu}\|\bar{\pi}(\xi))}
	.
\end{equation}

To evaluate the left hand side of this extremization condition,
I use Eq. (\ref{shiftreference}) and
the Gaussian approximation to write
\begin{equation}
	\sum_{i=1}^R w_i S(\pi_i(\xi)\|\bar{\pi}(\xi))
	\approx
	\frac{1}{2 C(\bar{\mu})} \mbox{var}(U)
	,
\end{equation}
where
\begin{equation}
	C(\bar{\mu})
	:=
	\langle \delta H(\xi) ; \delta H(\xi) \rangle_{\bar{\mu}}
\end{equation}
with
$\delta H(\xi):=H(\xi)-\bar{U}(\xi)$,
and
\begin{equation}
	\mbox{var}(U):=\sum_{i=1}^R w_i (U_i-\bar{U})^2
	.
\end{equation}
The latter two functions have the respective derivatives
\begin{equation}
	{\partial_\xi C(\bar{\mu})}
	=
	2 \langle \delta H(\xi) ; \delta(\partial_\xi H) \rangle_{\bar{\mu}}
\end{equation}
with 
$\delta(\partial_\xi H):={\partial_\xi H}-{\partial_\xi \bar{U}}$
and
\begin{equation}
	{\partial_\xi \mbox{var}(U)}
	=
	2 \, \mbox{cov}(U,{\partial_\xi U})
\end{equation}
with covariance
\begin{equation}
	\mbox{cov}(U,{\partial_\xi U})
	:=
	\sum_{i=1}^R w_i (U_i-\bar{U})
	({\partial_\xi U_i} - {\partial_\xi \bar{U}})
	.
\end{equation}
The right hand side of the extremization condition is given by
\begin{equation}
	{\partial_\xi S(\bar{\mu}\|\bar{\pi}(\xi))}
	=
	\bar{\beta} 
	(
	\langle 
	{\partial_\xi H} 
	\rangle_{\bar{\mu}} - 
	\langle 
	{\partial_\xi H} 
	\rangle_{\bar{\pi}(\xi)}
	)
	.
\end{equation}
Altogether, this yields the condition
\begin{eqnarray}
	&&
	\mbox{cov}(U,\partial_\xi U)
	- C(\bar{\mu})^{-1} \mbox{var}(U) \langle \delta H(\xi) ; \delta(\partial_\xi H) \rangle_{\bar{\mu}}
	\nonumber \\
	&&
	=
	\bar{\beta} C(\bar{\mu}) 
	(
	\langle 
	{\partial_\xi H} 
	\rangle_{\bar{\mu}} - 
	\langle 
	{\partial_\xi H} 
	\rangle_{\bar{\pi}(\xi)}
	)
	.
\label{mle_condition}
\end{eqnarray}

One particularly simple ansatz for the Hamiltonian  
is the linear form
\begin{equation}
	H(\xi)= - \sum_b \xi^b F_b
	,
\label{linear_ansatz}
\end{equation}
modulo some additive constant.
For the implementation of this ansatz it will be convenient to adopt a number of index conventions
in the style of general relativity:
Identical upper and lower indices are to be summed over;
the correlation matrix $C$ (Eq. (\ref{correlation})) and its inverse $C^{-1}$ lower or raise indices, respectively, akin to a metric tensor \cite{balian:physrep};
and the scalar product is defined as 
$x\cdot y:=x^a y_a=C_{ab}x^a y^b=(C^{-1})^{ab}x_a y_b$.
Furthermore, I define
the covariance matrix
\begin{equation}
	\Gamma_{ab}:=\sum_{i=1}^R w_i (f_a^i-\bar{f}_a)(f_b^i-\bar{f}_b)
\end{equation}
with $\bar{f}_b := \sum_i w_i f_b^i$,
its ``expectation value''
\begin{equation}
	\langle \Gamma\rangle_\xi
	:=
	\frac{\xi\cdot\Gamma\xi}{\xi\cdot\xi}
	,
\end{equation}
as well as
\begin{equation}
	\delta f_b(\xi):=\langle F_b\rangle_{\bar{\pi}(\xi)}-\bar{f}_b
\end{equation}
and $N:=\sum_i N_i$.
With these conventions and definitions
the asymptotic log-likelihood (\ref{asymptotic})
for the level of description ${\cal H}(\xi):=\mbox{span}\{1,H(\xi)\}$ reads
\begin{equation}
	L({\cal H}(\xi))
	\sim
	(N/2)
	[\langle\Gamma\rangle_{\xi} - \delta f(\xi) \cdot \delta f(\xi)]
	-
	(\Lambda/2)
	.
\label{xi_likelihood}
\end{equation}

If one is still uncertain as to whether the process in question has actually led to thermalization,
yet can already exclude the existence of other constants of the motion besides the Hamiltonian,
one must compare the log-likelihood of ${\cal H}(\xi)$ for all values of $\xi$
with the log-likelihood of ${\cal F}$,
i.e., the hypothesis that the data do not warrant any dimensional reduction at all.
The latter log-likelihood is given by
\begin{equation}
	L({\cal F})
	\sim
	(N/2)
	{\mbox{tr}(\Gamma)}
	- (\Lambda/2) {\dim \pi^\sigma_{\cal F}({\cal S})} 
	,
\label{perfect-fit_likelihood}
\end{equation}
where $\mbox{tr}(\Gamma):=\Gamma^a_a$.
The process may be considered ``thermalizing'' with Hamiltonian $H(\xi)$ iff
$L({\cal F})\ll L({\cal H}(\xi))$,
and hence
\begin{equation}
	[\mbox{tr}(\Gamma)
	-
	\langle\Gamma\rangle_{\xi}]
	+
	\delta f(\xi)\cdot \delta f(\xi)
	\ll
	(\Lambda/N)
	[\dim \pi^\sigma_{\cal F}({\cal S}) - 1]
	.
\label{thermal_condition}
\end{equation}

The most likely value of $\xi$ is determined by the 
maximum likelihood condition (\ref{mle_condition}),
which for the linear ansatz (\ref{linear_ansatz}) simplifies to
\begin{equation}
	(\delta_{\xi}\Gamma) {\xi}  
	=
	\bar{\beta} ({\xi}\cdot{\xi}) \delta f(\xi)
\label{mle_condition2}
\end{equation}
with matrix
$\delta_\xi \Gamma:=\Gamma - \langle \Gamma\rangle_\xi$.
In order to estimate $\bar{\beta}$, I consider
\begin{equation}
	\delta(\ln\bar{\pi}(\xi)) = \bar{\beta} \xi\cdot \delta F + \ln\sigma - \langle\ln\sigma\rangle_{\bar{\mu}}
	,
\end{equation}
where as before $\delta X:=X-\langle X\rangle_{\bar{\mu}}$.
In the typical case of a {uniform reference state} $\sigma$
the latter two terms cancel 
so that
\begin{equation}
	\bar{\beta}^2 \xi\cdot\xi
	=
	\langle \delta(\ln\bar{\pi}(\xi));\delta(\ln\bar{\pi}(\xi))\rangle_{\bar{\mu}}
	.
\end{equation}
The right hand side in turn may be approximated
to lowest order by
\begin{equation}
	\langle \delta(\ln\bar{\pi}(\xi));\delta(\ln\bar{\pi}(\xi))\rangle_{\bar{\mu}}
	\approx	
	\langle \delta(\ln\bar{\mu});\delta(\ln\bar{\mu})\rangle_{\bar{\mu}}
	.
\end{equation}

\section{\label{qubit}Example: Qubits}

In the following, I shall illustrate the general framework in the simple example of qubits,
which is tractable analytically.
In this example the $\{F_b\}$ are the Pauli operators,
and the parameter vector ${\xi}$ may be viewed as (parallel to) an effective magnetic field.
In the typical case of a uniform reference state $\sigma$
the expectation values of $F$ in the states $\bar{\mu}$ and $\bar{\pi}(\xi)$ are related linearly:
\begin{equation}
	\langle F\rangle_{\bar{\pi}(\xi)} = \frac{\xi\cdot\bar{f}}{\xi\cdot\xi} \xi
	.
\end{equation}
As a result, the maximum likelihood condition (\ref{mle_condition2}) becomes
\begin{equation}
	(\delta_{\xi}\Gamma) \xi
	=
	\frac{(\xi\cdot\bar{f})^2}{\xi \cdot \xi} \xi
	-
	(\xi\cdot\bar{f}) \bar{f}
	.
\label{mle_qubits}
\end{equation}
This condition no longer depends on $\bar{\beta}$,
and moreover, is invariant under rescaling of $\xi$.
Without loss of generality, therefore,
$\xi$ may be taken to be normalized, $\xi \cdot \xi=1$.

For qubits the covariance matrix $\Gamma$ is a $3\times 3$ matrix.
To simplify matters, I assume that it singles out one dominant direction $\gamma$, and is isotropic in the remaining two directions:
\begin{equation}
	\Gamma \xi
	=
	\Gamma_+ (\gamma\cdot\xi) \gamma + \Gamma_- P \xi
	,
\end{equation}
where the projector $P$ projects orthogonally (with respect to the scalar product used here)
onto the subspace complementary to $\gamma$,
and $\Gamma_+,\Gamma_-$ with $\Gamma_+>\Gamma_-$ are the respective eigenvalues.
The unit vectors
$\gamma$ and $\hat{f}$ 
(the unit vector pointing in the direction of $\bar{f}$)
then constitute the two preferred directions in the problem.
Symmetry dictates that the solution of the maximum likelihood condition (\ref{mle_qubits}) must lie in the subspace spanned by these two preferred directions, $\xi\in\mbox{span}\{\gamma,\hat{f}\}$.
In fact,
if $\gamma$ is aligned with $\hat{f}$,
the solution is $\xi=\gamma=\hat{f}$.
In case $\gamma$ and $\hat{f}$ are not aligned,
the solution will generally not coincide with either of the two.

In order to quantify how $\xi$ interpolates between $\gamma$ and $\hat{f}$ 
in case the two are not aligned,
I define a further unit vector
$\eta:\propto P\hat{f}$,
the normalized projection of $\hat{f}$ onto the subspace complementary to $\gamma$.
To lowest (first) order perturbation theory in $(\eta\cdot\hat{f})$,
i.e., for small misalignments,
the maximum likelihood condition (\ref{mle_qubits}) has the solution
\begin{equation}
	\eta\cdot\xi
	\approx
	\left[
	1 + \frac{\Gamma_+ - \Gamma_-}{\bar{f}\cdot\bar{f}}
	\right]^{-1}
	\eta\cdot\hat{f}
	\ ,\ 
	\gamma\cdot\xi \approx 1 - O((\eta\cdot\hat{f})^2)
	.
\label{interpol}
\end{equation}
This result illustrates nicely how the maximum likelihood algorithm interpolates between alignment with the center of mass ($\eta\cdot\xi=\eta\cdot\hat{f}$) and alignment with the covariance pattern ($\eta\cdot\xi=0$).
For a perfectly isotropic covariance pattern ($\Gamma_+=\Gamma_-$) the parameter vector is aligned with $\hat{f}$.
The more pronounced the anisotropy of the covariance pattern ($\Gamma_+\gg\Gamma_-$)
and the smaller the lever of the center of mass ($\bar{f}\cdot\bar{f}$ small),
the more $\xi$ tends to be aligned with $\gamma$.

Inserting the maximum likelihood solution into the formula (\ref{xi_likelihood}) for the log-likelihood yields,
to lowest order in perturbation theory,
\begin{eqnarray}
	\lefteqn{
	\max_\xi L({\cal H}(\xi))
	}
	\nonumber \\
	&\sim&
	\frac{N}{2}
	\left\{
	\Gamma_+
	-
	\left[
	\frac{1}{\bar{f}\cdot\bar{f}} + \frac{1}{\Gamma_+ - \Gamma_-}
	\right]^{-1}
	(\eta\cdot\hat{f})^2
	\right\}
	-\frac{\Lambda}{2} 
	.
\label{xi_mle}
\end{eqnarray}
The maximum likelihood solution satisfies
the thermalization condition (\ref{thermal_condition}) if and only if
\begin{equation}
	\Gamma_- \ll \frac{\Lambda}{N} 
	\ ,\ 
	\frac{\theta^2}{2}
	\ll
	\frac{\Lambda}{N}
	\left[
	\frac{1}{\bar{f}\cdot\bar{f}}
	+
	\frac{1}{\Gamma_+ - \Gamma_-}
	\right]
	,
\label{thermal}
\end{equation}
where $\theta$ is the tilting angle between $\gamma$ and $\hat{f}$,
$\sin \theta:=\eta\cdot\hat{f}$.

As a simple numerical example,
I consider data gleaned from multiple qubit samples of identical size $N_i=20,000$, and hence $\ln N_i\approx 10$.
I assume that the distribution of tomographic images has a width which in the dominant direction is
of comparable magnitude as the distance of the center of mass from the origin;
more specifically, that both are about $1/10$ of the radius of the Bloch sphere,
$\Gamma_+\approx \bar{f}\cdot\bar{f}\approx 1/100$.
In the other directions, the standard deviation of the tomographic images is assumed to be smaller by a factor $10$,
$\Gamma_- \approx \Gamma_+/100$.
The dominant direction $\gamma$ and the orientation $\hat{f}$ of the center of mass are not aligned;
rather, they are tilted against each other by an angle $\theta=\pi/16$.
This raises doubts about thermalization,
as the canonical curves of a qubit are straight lines through the origin of the Bloch sphere.
May the qubits nevertheless be considered thermalized?
In fact they may, as the standard deviation and the tilting angle still satisfy both
thermalization conditions in Eq. (\ref{thermal}).
Their most plausible Hamiltonian, parametrized as in the ansatz (\ref{linear_ansatz}), contains an effective magnetic field $\xi$ which (modulo rescaling) is given by Eq. (\ref{interpol}), 
and which in this example approximately bisects the angle between $\gamma$ and $\hat{f}$.

In the above example the preferred axis $\xi$ of the Hamiltonian is inferred from the data,
rather than given or conjectured from the outset.
This distinguishes this example from other inference tasks where
one weighs the hypothesis of some \textit{a priori} fixed axis against the hypothesis that no such preferred axis exists,
for instance when comparing Ising and Heisenberg models for an anisotropic ferromagnet
on the basis of a single sample \cite{PhysRevA.84.012101}.

\section{\label{conclusions}Conclusions}

In this paper
I focused not
on the theoretical question whether or not some system with a given Hamiltonian \textit{ought to} thermalize,
but on the practical question whether or not experimental data indicate that a system with hitherto unknown dynamics \textit{has} actually thermalized.
This issue never really arises for systems that are macroscopic.
Outside the macroscopic realm, however,
and with data pertaining to small samples composed of, say, a few hundred system copies only, it becomes a nontrivial statistical inference task.
I have laid out the appropriate statistical framework for assessing thermalization under such adverse conditions.

In case the data do support the hypothesis of thermalization, 
and provided there is no evidence for additional constants of the motion,
I have shown how the data can be used to estimate the system's unknown Hamiltonian.
Hamiltonian estimation is increasingly important in quantum technology,
as it is needed to assess and certify the proper functioning of quantum devices.
Since my estimation scheme is based on studying thermal properties rather than time evolution
and thus requires output data only,
it may constitute a viable alternative to conventional time-based approaches
especially in situations where initial states or time are difficult to control.

Aside from its practical relevance, 
the framework presented here is also of interest conceptually.
One example is a better understanding of the iterative dynamics of thermalization.
Whenever a physical system exhibits a hierarchy of time scales,
thermalization typically occurs in stages, on successively longer time scales.
For instance, a dense plasma, initially in the kinetic regime far from equilibrium,
might quickly equilibrate locally and thus enter the hydrodynamic regime,
but only much later reach global equilibrium \cite{Balian1987229,rau:physrep}.
Associated with these various stages are successively smaller levels of description;
in this particular example, 
first the Boltzmann level of description (all single-particle observables),
then the hydrodynamic level of description (local particle, energy and momentum density),
and finally the equilibrium level of description (total energy and particle number).
Thermalization is thus accompanied by a sequence of level contractions.
The framework developed here provides the quantitative criterion as to when exactly these level contractions are warranted.

I see four routes for further research.
First,
it will be important to test the mathematical framework developed here on real or simulated experimental data.
In principle, any experiment that probes only tiny samples of matter such as an array of atoms or the debris from a single high energy collision \cite{becattini:thermal} will lend itself to such an analysis.
Processing the data will likely require the use of suitable numerical techniques.

Second,
in the present paper I made a number of idealizing assumptions.
For instance, I assumed that the only source of experimental error is projection noise due to the finiteness of the samples,
whereas there is no error stemming from inaccuracies of the measurement devices.
Moreover, I took the tomographic measurement setup to be near optimal in the sense of the quantum Stein lemma.
In case the observables $\{F_b\}$ do not commute,
this may involve global measurements which are difficult to implement in practice.
In future work
I plan to investigate how the mathematical framework must be adapted when these assumptions are relaxed.

Third, 
on a more conceptual level,
I consider it worthwhile to generalize the mathematical framework in the following way.
While the approach laid out in the present paper aims to infer the most plausible level of description in a single step,
a different approach might split this into two distinct inference tasks:
first estimating the optimal \textit{dimension} of the level of description;
and then, given the dimension, its optimal \textit{orientation}.
In this alternative approach the first step involves an additional Occam factor,
and so in principle, might lead to other conclusions than the present approach.
It will be interesting to understand under which circumstances such divergent conclusions may arise, and why.

Finally,
also on the conceptual level,
the pivotal log-likelihood function $L({\cal G})$  which features in the statistical analysis
depends 
on a number of scaling parameters:
the total number $R$ of samples, their sizes $\{N_i\}$, the Gibbs manifold dimension $p$, and -- when calibrating against $L({\cal F})$ -- the number of different measurement setups. 
I propose to investigate in more detail how the log-likelihood scales with each of these parameters,
and whether any general conclusions can be drawn from this about the typicality of thermalization in different scaling regimes.


\begin{acknowledgments}
I thank 
Emmanuel Klinger for pertinent questions 
and
Carl Edward Rasmussen for directing me to the literature on factor analysis and principal component analysis.
\end{acknowledgments}


%

\end{document}